\documentclass[twocolumn,showpacs,prl,superscriptaddress,floatfix]{revtex4}

\usepackage{color}
\usepackage{tabularx}
\usepackage{epsfig}
\usepackage{amsmath}
\usepackage{amssymb}
\usepackage{graphicx}

\begin{document}

\title{Behavior of Ising Spin Glasses in a Magnetic Field}

\author{Thomas J\"org} 
\affiliation {LPTMS, UMR 8626 CNRS et Universit\'e Paris-Sud, 91405 Orsay 
CEDEX, France}
\affiliation{\'{E}quipe TAO - INRIA Futurs, 91405 Orsay CEDEX, 
France}

\author{Helmut G.~Katzgraber} 
\affiliation {Theoretische Physik, ETH Z\"urich, CH-8093 Z\"urich, 
Switzerland}

\author{Florent Krz\c{a}ka{\l}a} 
\affiliation {Laboratoire PCT, UMR Gulliver CNRS-ESPCI 7083, 10 rue 
Vauquelin, 75231 Paris, France}

\begin{abstract}
We study the existence of a spin-glass phase in a field using
Monte Carlo simulations performed along a nontrivial path in
the field--temperature plane that must cross any putative de
Almeida-Thouless instability line. The method is first tested
on the Ising spin glass on a Bethe lattice where the instability
line separating the spin glass from the paramagnetic state is also
computed analytically.  While the instability line is reproduced by our
simulations on the mean-field Bethe lattice, no such instability line
can be found numerically for the short-range three-dimensional model.
\end{abstract}


\maketitle

Since its proposal in the mid-70's, the Edwards and Anderson
Ising spin-glass Hamiltonian has become a source of inspiration in
statistical physics, especially in the context of mean field theory
\cite{edwards:75,sherrington:75,mezard:87,young:98}, and has been
applied to a wide variety of problems across scientific disciplines.
However, basic yet simple questions about the very nature of the
spin-glass state in (experimentally relevant) finite space dimensions
are still subject of controversy. The most prominent of such open
questions is the existence of spin-glass ordering in a magnetic field.

\begin{figure}
\includegraphics[width=0.8\columnwidth]{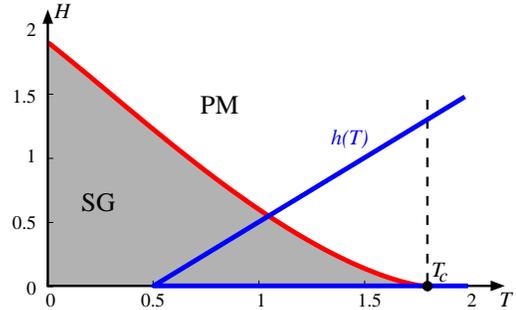}
\vspace*{-0.2cm}
\caption{(Color online)
Magnetic field $H$ versus temperature $T$ phase diagram of the Ising
spin glass on a mean-field random regular graph with connectivity $c =
6$ with Gaussian interactions computed using the cavity method. 
The paramagnetic (PM) phase is separated from the
spin-glass (SG) phase by the de Almeida-Thouless line (light red curve).
The diagonal (dark blue) line $h(T)$ represents the simulation path
followed in the Monte Carlo simulations. Because the path starts at
$T > T_c$ for $H = 0$ and then increases along a diagonal $h(T) =
r(T - T_{\rm min})$ in the $H$--$T$ plane above $T_{\rm min} \ll T_c$
up to a value beyond $T_c$ (dashed vertical line), an intersection
with a putative AT line is guaranteed.  The same approach is used
for the diluted 3D simulations where $T_c = 0.663(6)$ \cite{joerg:06}.
\label{fig:at}
\vspace{-0.8cm}
}
\end{figure}

The fully-connected mean-field version of the EA model, called the
Sherrington-Kirkpatrick (SK) model \cite{sherrington:75}, was solved
using the replica method by Parisi \cite{mezard:87,young:98};
the obtained free energy recently proven to be rigorously
exact by Talagrand \cite{talagrand:06}.  In this model,
the low-temperature spin-glass phase is characterized by a
complex free energy landscape made of many different valleys.
This phase also exists for low externally-applied magnetic fields
and the so-called de Almeida-Thouless (AT) instability line
\cite{almeida:78} separates the spin-glass from the paramagnetic
phase at finite fields/temperatures.  A similar scenario arises in
the Bethe lattice approximation (see Fig.~\ref{fig:at}). However,
within the more phenomenological description of spin glasses
known as the droplet picture \cite{fisher:86} any infinitesimal
field destroys the spin-glass order, in stark disagreement with
the aforementioned mean-field description.  Numerical evidence
favoring the absence of a spin-glass state in a field for short-range
spin glasses below the upper critical dimension have become stronger
\cite{migliorini:98,houdayer:99,young:04,krzakala:05a,jonsson:05-ea,katzgraber:05c,sasaki:07b-ea},
although different opinions remain
\cite{ciria:93-ea,marinari:98d-ea,krzakala:01-ea,parisi:07}.

Capitalizing on the success of studying the susceptibility and the
finite-size correlation length \cite{palassini:99b,ballesteros:00-ea}
to probe the spin-glass phase, we study the problem using a novel
numerical approach backed up with analytic calculations on the Bethe
lattice.  On the numerical side, we use a multi-spin coded version
of exchange (parallel tempering) Monte Carlo \cite{hukushima:96}
with a new twist where the replicas ``live'' in the $H$--$T$ plane
along a nontrivial path that {\em guarantees} a crossing with a
potential AT line (see Fig.~\ref{fig:at}) and that in contrast
to Refs.~\cite{billoire:03,billoire:03b} has both ends in the
high temperature region where decorrelation is fast.  We solve the
thermalization problems on diagonal paths found in Ref.~\cite{young:04}
by studying the link-diluted version of the model \cite{joerg:06}
where the cluster moves introduced in Refs.~\cite{joerg:05} and
\cite{joerg:06} can be used. In this model, the cluster updates allow
the thermalization times to be decreased by a factor of at least
$10^3$, allowing us to probe relatively large system sizes down to
very low temperatures and large magnetic fields.  We first demonstrate
the efficiency of our strategy by applying it on the model defined
on a regular random graph---which corresponds to a Bethe lattice
in this context---for which we compute the AT line analytically,
generalizing the original result by de Almeida and Thouless for the
SK model \cite{almeida:78}.  The analytical and numerical results
for the Bethe lattice agree to high precision. However, results on
the 3D model show no sign of an AT line. The fact that we {\em do}
observe an AT line for the mean-field Bethe lattice, exactly where it
is predicted, and {\em do not} for the short-range 3D Ising spin glass
is convincing evidence that the phase diagram is indeed {\em trivial}
and there is no spin-glass state in a field in three dimensions (3D).

\paragraph*{Models ---}
\label{sec:model}

The spin-glass Hamiltonian is given by
\begin{equation}
{\mathcal H} = - \sum_{i, j} J_{ij} S_i S_j - H \sum_i S_i .
\label{eq:ham}
\end{equation}
The Ising spins $S_i \in \{\pm 1\}$ have nearest-neighbor
interactions. We study the mean-field case where $N$ spins lie on
the vertices of a regular random graph, and the EA model on a cubic
lattice of size $N = L^3$ with periodic boundary conditions, where the
interactions $J_{ij} \in \{-1,0,1\}$ are chosen from a link-diluted
bimodal distribution with a link occupation probability of $45 \%$
\cite{joerg:06}.

\paragraph*{Theoretical predictions ---}
\label{sec:mf}

Following recent progress in the study of finite-connectivity
mean-field systems, it is now possible to study (quite) precisely
spin-glass models on Bethe lattices using the cavity method
\cite{mezard:01}.  Within this formalism, first steps in computing
the phase diagram in the field--temperature plane have been achieved
(see Refs.~\cite{pagnani:03-ea} and \cite{krzakala:05a}). In order to
determine the AT line, one needs to compute the onset of divergence of
the spin-glass susceptibility $\chi$ in the paramagnetic phase in which
case, thanks to the finite correlation length, the model is equivalent
to a spin glass on an infinite tree with random boundary conditions
\cite{mezard:01} (this is reminiscent of the exact analysis presented
in Ref.~\cite{carlson:88-ea}).  We thus consider here the simplest
Bethe-Peierls cavity approach for the problem and compute the AT
line in this context.  We refer the reader to Refs.~\cite{mezard:01},
\cite{pagnani:03-ea}, \cite{krzakala:05a}, and \cite{martin:05-ea} for
a description of the method. For $H = 0$, we recover the well-known
result first found by Thouless \cite{thouless:86}: $k[\tanh(\beta_c
J)^2]_{\rm av} = 1$, where $\beta_c = 1/T_c$, $k = c - 1$ ($c$
the connectivity), and $[\cdots]_{\rm av}$ is a disorder average.
For finite external field, one has to perform a numerical solution
of the cavity equations to solve the cavity recursion and to compute
the point where the susceptibility diverges \cite{martin:05-ea}.

Using this approach, we compute points along the AT line for
different values of the field $H$ with Gaussian and bimodal disorder
and connectivities $c = 3$ and $6$.  We have also checked that the
large-connectivity limit of our computation yields the SK result.
We find that, close to $T_c$, the data scale as $h_{\rm AT}(T) \sim
(T_c - T)^{3/2}$---as first predicted in Ref.~\cite{thouless:86} and
checked in Ref.~\cite{pagnani:03-ea}---and the instability line is then
approximately linear close to zero temperature \cite{comment:pagnani}.
A fit to $h_{\rm AT}(T)= a_1 (T_c-T)^{3/2} \exp(a_2 T +a_3 T^2)$
gives very accurate results whose precision is within the error
of our numerical evaluation of the cavity recursion. For $c=6$, we
obtain $a_1 = 0.786$ ($0.875$), $a_2 = 0.111$ ($0.221$), and $a_3 =
-0.054$ ($-0.127$) and $T_c = 1.807$ ($2.078$) for Gaussian (bimodal)
distributed disorder. For $c=3$, we find $a_1 = 0.785$ ($0.827$), $a_2
= 0.251$ ($0.413$), and $a_3 = 0.117$ ($ 0.083$) and $T_c = 0.748$
($1.135$). The Gaussian case with $c=6$ is shown in Fig.~\ref{fig:at}
(light red curve).

The presence of the instability line in finite-dimensional systems
has been criticized before, especially in the context of the droplet
model \cite{fisher:86}.  Other pictures were proposed where no
such line is present \cite{krzakala:00,palassini:00,newman:03}.
It has been also suggested that the line disappears
below the upper critical dimension $d_{\rm u} = 6$
\cite{bray:80b,temesvari:02-ea,pimentel:02-ea,katzgraber:05c},
and even more complex scenarios are possible
\cite{temesvari:02a,temesvari:06,temesvari:07}.  We now attempt to
clarify this issue in the 3D case.

\paragraph*{Numerical method ---}
\label{sec:num}

The simulations are performed using exchange Monte Carlo (EMC)
\cite{hukushima:96}. Traditionally, the field is fixed at a constant
value and replicas at different temperatures perform a Markov chain
in temperature space.  There have been alternate approaches where
the temperature $T$ is fixed to $T < T_{\rm c}$ and the replicas
perform a Markov chain in field space. This method has not proven
to be efficient because tunneling across $H = 0$ requires special
moves \cite{billoire:03,billoire:03b} and especially because in
this case no replica has the chance to reach the zero-field high-$T$
phase where relaxation is fast.  In order to solve this problem, we
propose an EMC method in the $H$--$T$ plane (see Fig.~\ref{fig:at})
at which part of the replicas have labels $(T,H = 0)$ for temperatures
values in the range $T_{\rm min} \le T \le T_{\rm max}$ with $T_{\rm
max} \gg T_{\rm c}$ (in 3D $T_{\rm min} = 0.30$). We also couple
the replicas in the EMC scheme to a second set of replicas along a
diagonal path in the $H$--$T$ plane with labels $(T,H)$ and $h(T) = r
(T - T_{\rm min})$, $r$ constant.  It is important not to choose the
slope $r$ of $h(T)$ too steep because EMC is least efficient for large
fields \cite{moreno:03-ea,billoire:03,billoire:03b}. In addition,
$T_{\rm max}(H > 0) > T_c(H = 0)$ has to be chosen to ensure that
the a potential AT line is crossed.  The method has the advantage to
deliver data at zero (horizontal blue path in Fig.~\ref{fig:at}),
as well as finite field (diagonal blue path in Fig.~\ref{fig:at}).
Finally, it is advantageous to choose the slope of the path in the
$H$--$T$ plane such that it crosses the phase boundary orthogonally
(cleaner signal of the transition).

We test equilibration by a logarithmic binning of the data until
the data in the last three bins agree within error bars.  To access
relatively large system sizes in 3D, we use a multi-spin coded version
of the program that updates 32 copies of the system in parallel. In
order to obtain thermalization in reasonable time, it is necessary
to apply the cluster algorithm introduced in Refs.~\cite{joerg:06}
and \cite{joerg:05}. The speedup obtained by this approach is of three
orders of magnitude over conventional approaches, even for the smallest
sizes simulated. Note that the speedup increases with increasing
system size; our simulations would not have been possible otherwise.
Simulation parameters are listed in Table \ref{tab:simparams}.

\begin{table}
\caption{
Simulation parameters: $N_{\rm sa}$ is the number of samples,
$N_{\rm sw}$ is the number of Monte Carlo sweeps for one 
sample, $T_{\rm min}$ is the lowest temperature simulated, $H_{\rm
max}$ is the maximum field studied and $N_{\rm r}$ is the number
of replicas used in the EMC method. $h_{\rm BL}(T) = 1.0 (T-0.5)$
for the Bethe lattice (top set ); in 3D $h_{\rm 3D}(T) = 0.7 (T-0.3)$ 
(bottom set).
\label{tab:simparams}
}
{\footnotesize
\begin{tabular*}{\columnwidth}{@{\extracolsep{\fill}} r r r c c c }
\hline 
\hline 
$N$  &  $N_{\rm sa}$  & $N_{\rm sw}$ & $T_{\rm min}$ & $H_{\rm max}$ & $N_{\rm r}$\\ 
\hline
$64$   & $12834$ & $100000$  & $0.5$ & $2.0$ & $25$ \\
$128$  & $12689$ & $100000$  & $0.5$ & $2.0$ & $25$ \\
$256$  & $5107$  & $200000$  & $0.5$ & $2.0$ & $25$ \\
$512$  & $2546$  & $1000000$ & $0.5$ & $2.0$ & $49$ \\
\hline
\hline
$L\phantom{^*}$  &  $N_{\rm sa}$  & $N_{\rm sw}$ & $T_{\rm min}$ & $H_{\rm max}$ & $N_{r}$\\ 
\hline
$4\phantom{^*}$  & $20000$ & $200000$  & $0.3$ & $0.49$ & $47$ \\
$5\phantom{^*}$  & $20000$ & $400000$  & $0.3$ & $0.49$ & $47$ \\
$6\phantom{^*}$  & $20000$ & $400000$  & $0.3$ & $0.49$ & $47$ \\
$8\phantom{^*}$  & $10528$ & $1000000$ & $0.3$ & $0.49$ & $51$ \\
$10\phantom{^*}$ & $6080$  & $2000000$ & $0.3$ & $0.49$ & $51$ \\
$12\phantom{^*}$ & $2944$  & $4000000$ & $0.3$ & $0.49$ & $51$ \\
\hline
\hline
\end{tabular*}
}
\end{table}

\paragraph*{Results on the Bethe lattice ---}
\label{sec:results_mf}

\begin{figure}
\includegraphics[width=0.95\columnwidth]{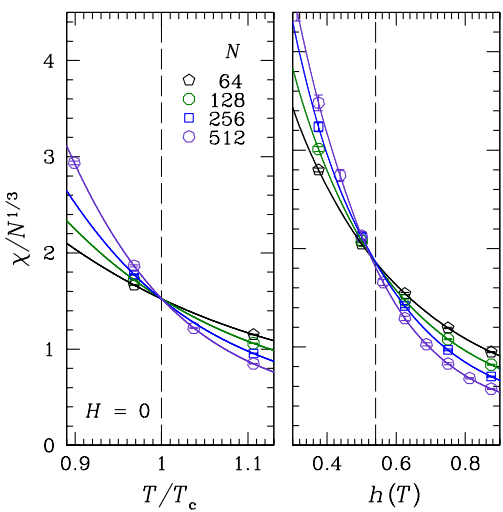}
\vspace*{-1.2cm}
\caption{(Color online)
Rescaled susceptibility $\chi/N^{1/3}$ from Monte Carlo data of the
regular random graph model with Gaussian disorder and connectivity $c
= 6$. Left: Data for zero field cross at the analytically predicted
critical temperature $T_{\rm c} = 1.807$.  Right: Data along $h_{\rm BL}(T)
= 1.0 (T - 0.5)$.  The crossing point agrees very well with the
values in Fig.~\ref{fig:at}.
\label{fig:mean_field}
}
\end{figure}

We have computed the connected spin-glass susceptibility using $\chi
= N \left[\langle q^2\rangle_T-\langle q \rangle_T^2 \right]_{av}$
where $q = 1/N \sum_{i=1}^N q_i$ with $q_i=S^{1}_iS^{2}_i$ being
the overlap at site $i$ between two independent replicas of the
system. Here $\langle \cdots \rangle_T$ represents a thermal average.
The susceptibility of the mean-field model has a finite-size
scaling form $\chi(T,H) = N^{1/3} \widetilde{G}\left(N^{1/3}[T -
T_{\rm c}(H)]\right)$ \cite{billoire:03,billoire:03b} hence data for
different $N$ should cross at $T_c$ when plotted as $\chi/N^{1/3}$
versus $T$. This is shown in Fig.~\ref{fig:mean_field} for $c=6$ and a
Gaussian distribution of the interactions: The left panel is at zero
field, whereas the right panel is along the diagonal path $h_{\rm
BL}(T) = 1.0(T - 0.5)$. In both cases, the data cross in agreement
with the analytical results (Fig.~\ref{fig:at}). Hence our numerical
approach allows for a precise detection and location of any putative
AT line.

\paragraph*{Results in three dimensions ---}
\label{sec:results_3d}

In the finite-dimensional system the scaling of the susceptibility
requires an additional parameter $\eta$ whose putative in-field value 
is unknown, as opposed to the mean-field case. We thus prefer to
study the transition via the scaling of the finite-size two-point
correlation length \cite{palassini:99b,ballesteros:00-ea} given by
\begin{equation}
\xi_L = \frac{1}{2 \sin (|{\bf k}_\mathrm{min}|/2)}
\left[\frac{\chi(0)}{\chi({\bf k}_\mathrm{min})} -
1\right]^{1/2} ,
\label{eq:xiL}
\end{equation}
where ${\bf k}_\mathrm{min} = (2\pi/L,0,0)$ is the smallest nonzero
wave vector and $\chi({\bf k})$ the wave-vector-dependent spin-glass
susceptibility defined as
\begin{equation}
  \chi(\mathbf{k}) = \frac{1}{N} \sum_{i, j} \left[
    \langle q_i q_j\rangle_T - \langle q_i \rangle_T \langle q_j\rangle_T
  \right]_{\rm av} e^{i\mathbf{k}\cdot(\mathbf{R}_i -
    \mathbf{R}_j)} .
\label{eq:chisg}
\end{equation}
Note that we used a slightly different definition of the susceptibility
than in Ref.~\cite{young:04}. However, the two expressions are
equivalent, except that the definition used here requires only
two replicas (unlike the four needed in Ref.~\cite{young:04})
thus saving computational resources.  The correlation length
divided by the system size $L$ has the scaling relation $\xi_L/L =
\widetilde{X}\left(L^{1/\nu}[T - T_{\rm c}(H)]\right)$, where $\nu$
is the usual critical exponent.  When $T = T_{\rm c}(H)$ data for
different $L$ cross signaling the existence of a transition.

\begin{figure}
\includegraphics[width=0.95\columnwidth]{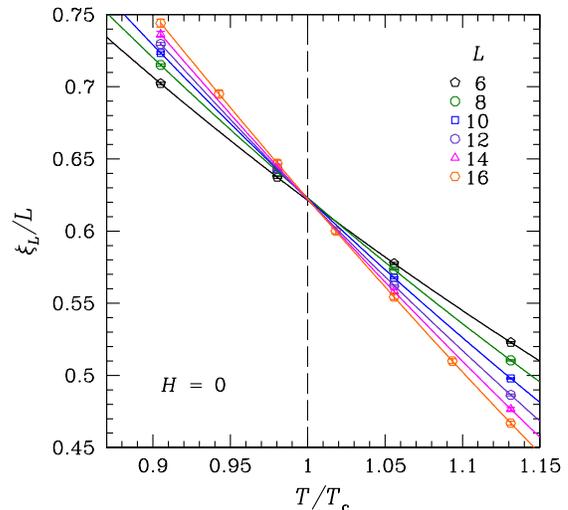}
\vspace*{-1.3cm}
\caption{(Color online)
Finite-size correlation length as a function of $T/T_{\rm c}$ for the
3D Ising spin glass at zero field with bimodal interactions. The data
cleanly cross at $T_{\rm c} = 0.663(6)$.
\label{fig:xiL_nofield}
}
\vspace*{-0.5cm}
\end{figure}

\begin{figure}
\includegraphics[width=0.95\columnwidth]{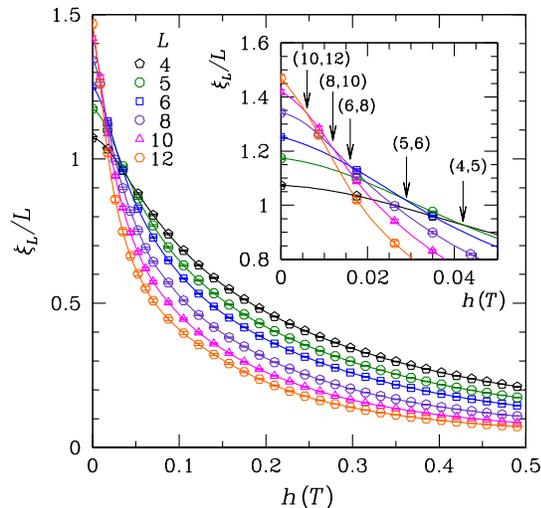}
\vspace*{-1.3cm}
\caption{(Color online)
Finite-size correlation length as a function of $h_{\rm 3D}(T) = 0.7 (T -
0.3)$ (finite-field branch in Fig.~\ref{fig:at}) for the 3D Ising
spin glass. The apparent crossing of the data wanders towards zero
for increasing system size (see inset for a zoom into the low-field
region), i.e., there is no sign of a transition for $H > 0$ (arrows
mark crossings between $L$-pairs).
\label{fig:xiL_field}
}
\vspace*{-0.89cm}
\end{figure}

Data for the 3D model in zero field are shown in
Fig.~\ref{fig:xiL_nofield}. A clear crossing point at $T_{\rm c} =
0.663(6)$ is observed, thus signaling the presence of a spin-glass
state for system sizes up to $L = 16$ and in agreement with previous
results \cite{joerg:06}.  In contrast, data along the finite-field
branch [$h_{\rm 3D}(T) = 0.7 (T - 0.3)$] (see Fig.~\ref{fig:at}) show
no sign of a transition (see Fig.~\ref{fig:xiL_field}). The inset
shows a zoom of the data for small fields where the crossing points
(arrows) between the different lines wander towards zero for increasing
$L$. This shows that at temperature $T=0.3$, the instability is below
$h<0.006$.  This should be compared with the mean field case where,
e.g., for $c=3$ with Gaussian couplings (having a similar $T_c$ as the
3D model), one observes $h(0.3)\approx0.26$, i.e., a value $40$ times
larger. Performing a one-parameter scaling with fixed volume ratio
$h_c(L) \sim h_c^\infty + b/L^2$ gives $h_c^\infty = -0.00004(14)$,
i.e., compatible with zero with a $Q$-factor of 94.9\%. Fixing
$h_c^\infty = 0$ gives $Q = 99.7$\%; i.e., the data suggest the
absence of a spin-glass state in a field.

\paragraph*{Summary and discussion ---}
\label{sec:conclusions}

We have presented calculations of the AT line of a spin glass on
a Bethe lattice using the cavity method. These results for the
mean-field model on the Bethe lattice allowed the validation of our
Monte Carlo simulations where we have observed an AT line close to
its predicted value. This is in stark contrast to the 3D Ising spin
glass with bimodally-distributed bonds where data for a wide range
of fields and temperatures clearly show a lack of ordering in a field.

It would be interesting to study higher-dimensional systems using
the method presented here to verify whether or not an AT line is
observed above the upper critical dimension \cite{katzgraber:05c}
and we hope our results will also spark new theoretical developments
in this direction.

\begin{acknowledgments}

We thank A.~P.~Young for helpful discussions.  The simulations
have been performed on the ETH Z\"urich Hreidar cluster.
H.G.K.~acknowledges support from the Swiss National Science Foundation
under Grant No.~PP002-114713. T.J.~acknowledges support from EEC's HPP
HPRN-CT-2002-00307 (DYGLAGEMEM) and FP6 IST contracts under IST-034952
(GENNETEC).

\end{acknowledgments}

\vspace{-0.6cm}
\bibliography{refs,comments}

\begin{thebibliography}{41}
\bibitem[{com()}]{comment:pagnani}
\bibinfo{note}{Our results contradict the results of Ref.~\cite{pagnani:03-ea}
  because there too small systems were used. Our results are recovered for
  larger system sizes.}
\end{thebibliography}

\end{document}